\pdfoutput=1

\documentclass[preprint]{aastex}
\usepackage{natbib}
\usepackage[T1]{fontenc}
\usepackage{ulem}
\usepackage{lscape}
\usepackage{longtable}
\usepackage{amsmath,amssymb} 
\usepackage{colortbl} 
\usepackage{rotating}
\definecolor{lightgold}{rgb}{1,0.93,0.55}

\shorttitle{IceCube Time-Dependent Searches from Binary Systems}
\shortauthors{R.~Abbasi et al.}

\usepackage[utf8]{inputenc}

\newcommand{\cygxo}{Cygnus X-1}
\newcommand{\cygxt}{Cygnus X-3}
\newcommand{\gro}{GRO~J0422+32}
\newcommand{\grs}{GRS~1915+105}
\newcommand{\hess}{HESS~J0632+057}
\newcommand{\lsi}{LS~I~+61$^\circ$~303}
\newcommand{\ls}{LS~5039}
\newcommand{\psr}{PSR~B1259-63}
\newcommand{\ssf}{SS~433}
\newcommand{\xte}{XTE~J1118+480}

\begin{document}

\title{Searches for periodic neutrino emission from binary systems with 22 and 40 strings of IceCube}

\author{
IceCube Collaboration:
R.~Abbasi\altaffilmark{1},
Y.~Abdou\altaffilmark{2},
T.~Abu-Zayyad\altaffilmark{3},
M.~Ackermann\altaffilmark{4},
J.~Adams\altaffilmark{5},
J.~A.~Aguilar\altaffilmark{1},
M.~Ahlers\altaffilmark{6},
M.~M.~Allen\altaffilmark{7},
D.~Altmann\altaffilmark{8},
K.~Andeen\altaffilmark{1,9}
% ...
}
\altaffiltext{1}{Dept.~of Physics, University of Wisconsin, Madison, WI 53706, USA}
\altaffiltext{2}{Dept.~of Physics and Astronomy, University of Gent, B-9000 Gent, Belgium}
\altaffiltext{3}{Dept.~of Physics, University of Wisconsin, River Falls, WI 54022, USA}
\altaffiltext{4}{DESY, D-15735 Zeuthen, Germany}
\altaffiltext{5}{Dept.~of Physics and Astronomy, University of Canterbury, Private Bag 4800, Christchurch, New Zealand}
\altaffiltext{6}{Dept.~of Physics, University of Oxford, 1 Keble Road, Oxford OX1 3NP, UK}
\altaffiltext{7}{Dept.~of Physics, Pennsylvania State University, University Park, PA 16802, USA}
\altaffiltext{8}{III. Physikalisches Institut, RWTH Aachen University, D-52056 Aachen, Germany}
\altaffiltext{9}{now at Dept. of Physics and Astronomy, Rutgers University, Piscataway, NJ 08854, USA}
% ...


\begin{abstract}

In this paper we present the results of searches for periodic neutrino
emission from a catalog of binary systems.  Such modulation, observed in the photon
flux, would be caused by the geometry of these systems.  In the analysis, 
the period is fixed by these photon observations, while the phase and 
duration of the neutrino emission are treated as free parameters to be fit
with the data. If the emission occurs during $\sim 20$\% or less of the total
period, this analysis achieves better sensitivity than a time-integrated
analysis. We use the IceCube data taken from May 31, 2007 to 
April 5, 2008 with its 22-string configuration, and from April 5, 2008
to May 20, 2009 with its 40-string configuration. 
No evidence for neutrino emission
is found, with the strongest excess occurring for \cygxt\ 
at $2.1\,\sigma$ significance after accounting for trials.  
Neutrino flux upper limits for both periodic
and time-integrated emission are provided.

\end{abstract}

\section{Introduction}
\label{sec:intro}

Recently, the observation of Very High Energy (VHE) $\gamma$-ray
emission from several high mass X-ray binaries has established a new
subclass of VHE or ``$\gamma$-ray-loud'' binaries. 
While much of the evidence from multi-wavelength observations favors leptonic emission, it is likely that a hadronic component is also accelerated in the jets of these
binary systems. 
These binary systems are powerful accelerators with jets capable of accelerating cosmic rays and can have a role as PeVatron accelerators of galactic cosmic rays.
The observation of neutrino emission would be clear evidence for
the presence of a hadronic component in the outflow of these sources. 
Four such binary systems, \psr, \ls, \hess, and \lsi, have been identified as persistent TeV
$\gamma$-ray emitters \citep{Aharonian:2005bv,Aharonian:2005eb,Aharonian:2007nh,Albert23062006}, while \cygxo\ is a
possible candidate.
The binary pulsar system PSR B1259-63 was also recently discovered to have periodic
emission in GeV photons \citep{Abdo:2011dj}. \psr\ 
is formed by
a B2Ve star orbited by a young $48\mathrm{\,ms}$ pulsar
\citep{1997ApJ...477..439T} both exhibiting a
strong wind. As observed in \citep{Aharonian:2005bv}, its VHE emission could come from Inverse Compton
scattering on shock-accelerated leptons from the interaction zone
between the pulsar and wind from the star though a hadronic interpretation cannot be excluded. 
On the other hand, the driving factor of the VHE emission in \cygxo, most probably a
black hole orbiting a super-giant O9.7 star \citep{2005MNRAS.358..851Z},
could be the interaction of the black hole with the strong stellar wind of the star.
However, there has been no other evidence for steady VHE 
emission of \cygxo, though a VHE flare of about $1\mathrm{\,h}$
was observed \citep{Albert:2007mf}.
\hess\, was recently seen to have a periodic modulation in X-rays by Swift,
with heightened TeV emission coincident with the X-ray maximum \citep{Bongiorno:2011vx}.

\lsi\ remains a mystery even after four decades of observations over a wide range of wavelengths, from radio \citep{0004-637X-575-1-427}, soft and hard X-ray \citep{2041-8205-719-1-L104, 0004-637X-733-2-89, MNR:MNR17152, 2006AnA...459..901S, 2000ApJ...528..454H}, GeV \citep{Abdo:2009pw} and TeV photons \citep{Albert:2009lsi,Acciari:2009gg}. The best measurement of its period, $P_1 = 26.4960 \pm 0.0028\,$d, comes from radio data (\citet{0004-637X-575-1-427} and references therein) with the orbital zero phase taken by convention at JD~2443366.775 \citep{1978Natur.272..704G}, but the same modulation has also been detected in other wavelengths, notably in the GeV/TeV band emission
\citep{Abdo:2009pw,Albert:2009lsi}. 
Together with LS 5039, discovered in the TeV $\gamma$ band 
\citep{Aharonian:2005bv}, LS I +61 303 lacks a strong evidence supporting the black hole or neutron star nature of the compact object. 
This prevents clear classification of them as microquasars or pulsar systems
(\citet{Paredes:2011vh,2000Sci...288.2340P}; and references therein). A  discussion of the different theoretical models for these systems is presented in \citet{BoschRamon:2008su}.
As with all of the binary systems above, the detection
of multi-TeV neutrinos would complement the VHE photon observations and 
 unequivocally prove the existence of hadronic acceleration.

The analysis in this paper has been performed using a likelihood method in which 
the underlying hypothesis is that the neutrino emission is periodically modulated
due to the geometry of the X-ray binary systems. Neutrinos would be produced by a 
beam of hadrons accelerated by the compact object and interacting 
with the matter of the massive star and its atmosphere. The periodic modulation
would be connected to the orbital motion of the system. This modulation is observed in photons
from radio to X-ray, and in the VHE band.  
The analysis is designed to incorporate only minimal assumptions regarding the neutrino emission.  The period is fixed to that observed in an electromagnetic band, while the phase and duration of neutrino emission are free parameters and not constrained to match the photon emission.   This is to account for the fact that
photons can be absorbed when the accelerator is behind the large star of the binary system 
while neutrino production can be enhanced if enough
matter is crossed.
The neutrino energy spectrum is fit with a simple power law with the index also a free parameter. 

The paper is organized as follows.
In Sec.~\ref{sec:IceCube} we describe the IceCube observatory and the data 
taken with two detector configurations  
\citep{IceCubeFirstYearPerformance_Achterberg:2006md}. The analysis method is described 
in Sec.~\ref{sec:method}, and the expected sensitivity and discovery potential are shown.  To avoid bias, the search has been performed in a blind fashion by
defining cuts before looking at the true times (equivalently, the right
ascension values) of the final event sample.
In Sec.~\ref{sec:results} we present the results of the search performed on a catalog of
seven galactic binary stars in the Northern sky.
The selected objects are considered as
microquasars in \citet{Distefano:2002qw}, where their expected emission
of neutrinos is calculated. While that paper is not specifically about the periodic emission from these sources, nonetheless
the objects considered there are promising neutrino emitters for which radio
observations allow identification of jet parameters such as the Lorentz
factor and the luminosity of the jet. All of the sources considered are located in the Northern Hemisphere, 
where IceCube is most sensitive \citep{Jon}.

\section{The IceCube data}
\label{sec:IceCube}
Construction of the IceCube Neutrino Observatory started with a first
string installed in the austral summer of 2005/2006
\citep{IceCubeFirstYearPerformance_Achterberg:2006md} and was
completed in December 2010. It is composed of an array of
86 strings with a total of 5160 Digital Optical Modules (DOMs)
instrumented between a depth of 1,450 and 2,450 m. The deep ice detector
is complemented at the surface by the extensive air shower array IceTop. 
Each DOM consists of a 25 cm diameter Hamamatsu photomultiplier
\citep{Abbasi:2010vc} and electronics to digitize the PMT output voltage
\citep{IceCubeDAQ_:2008ym}, all in a spherical, pressure-resistant glass housing.
IceCube observes the Cherenkov photons emitted by relativistic charged particles produced in high-energy neutrino 
interactions. The PMT signals are digitized and the charge  and arrival time of photons measured. The data taking and performances of the detector during the two seasons analyzed here are described in \cite{Abbasi:2009iv} and \cite{Jon}.
The main background comes from down-going muons due to cosmic ray interactions in the 
atmosphere above the detector: in the 40-string
configuration these trigger the detector at a rate of about 950~Hz and in the 22-string
configuration at about 350 Hz. 
During the austral summer the atmosphere above the South Pole gets warmer and thinner and the probability of pions generated in cosmic ray air showers to decay rather than interact increases \citep{Tilav:2010hj}, causing the trigger rate to vary by about $\pm10\%$.
A series of event selections and higher-level event reconstructions are applied to remove these downward-going events, while retaining upward-going tracks from muons induced by neutrinos which crossed the Earth.
At the final level of analysis, this remaining background of upward-going atmospheric neutrinos comes from many different directions on the other side of the Earth. Temperature effects average over a wide terrestrial region and the seasonal modulation is only a few percent. This variation has a period of one year, much longer than any period considered in this search.

The searches presented here used two data samples. The data taken with
the 22-string configuration have a livetime of 275.7
days, 89\% of the operation period from May 31,
2007 to April 5, 2008, or Modified Julian Date (MJD) 54251-54561. The sample
is described in \citet{Abbasi:2009iv}, and consists of 5114 events,
which are mostly neutrino induced upward-going muons with declinations from
-5$^{\circ}$ to +85$^{\circ}$. The deadtime 
is mainly due to test and calibration runs during and after the
construction season. 
The livetime of the 40-string data used in analysis
is 375.7$\mathrm{\,d}$ which is 92\% of the nominal operation period
from April 5, 2008 to May 20, 2009 (MJD
54561-54971). The handling and processing of the data to
obtain the final neutrino candidate event sample are fully described
in \citet{Jon}. The final 40-string sample contains 36,900 atmospheric
neutrino and muon events distributed over the whole sky, of which
14,121 events are upward-going (the rest are downward-going events from the Southern Sky, used for neutrino source searches strictly above PeV energies). 
The median angular resolution for the final sample for energies greater than 10 TeV is $< 1^{\circ}$.
The energy of each event is estimated using the density of photons along the muon track due to stochastic energy losses of pair production, bremsstrahlung and
photonuclear interactions which dominate over ionization losses for muons above 1 TeV. 
The energy resolution is about 0.3 in $\log_{10}$ of the muon energy in the detector between 10 TeV and $10^5$ TeV.
The estimated muon energy is a lower bound on the primary neutrino energy, since for interactions that occur outside the detector the muon loses energy over an unknown distance before reaching the detector.
(Energy distributions used internally within the analysis therefore refer to the observable muon energies.)
The muon neutrino flux upper limits at 90\% CL for
time-integrated searches (depending on declination) are between
$E^{2} dN/dE \sim 3 - 20 \times 10^{-12} \mathrm{\,TeV\,cm^{-2}\,s^{-1}}$ in the northern sky where the sources considered in this paper are located.

The 22 and 40-string data samples used in this paper were also used to look for bursting 
neutrino sources in \citet{mike} where the stability of the 
data taking is discussed in detail. 
Azimuthal geometry effects of the 22 and 40-string IceCube detectors (due to the fact that they are more elongated in 
one direction than in others)
and the rotation of the Earth interfere constructively for source periods
that match to multiples of a half sidereal day, which is not the case for any of
the source periods tested.

The limits in this paper were produced assuming a flux of
only muon neutrinos and antineutrinos at the Earth with simulated energies from $10^{8}$ to $10^{19}$ eV. For standard neutrino oscillations over astronomical distances \citep{Athar}, equal fluxes of all neutrino flavors at the Earth are expected from a source producing
neutrinos via pion decay with a ratio of  $\nu_{e} : \nu_{\mu} : \nu_{\tau} = 1:2:0$.
For the assumption of equal fluxes of muon and tau neutrinos at the
Earth, the resulting upper limits on the sum of both fluxes are about
1.7 times higher than if only muon neutrinos are
considered \citep{Jon}. This sets better limits than the expected factor of two due to oscillations
if no tau neutrinos were detectable. 
This is due to the tau decay channel into muons with a branching ratio of 17.7\%. In addition to this, 
tau leptons with energy greater than several PeV that may travel far
enough to be reconstructed as tracks in IceCube before decaying. 
For an $E^{-2}$ neutrino spectrum, the contribution due to the detectable tau
neutrino flux for sources at the horizon is 10\% and rising to 15\% for
sources in the Northern hemisphere.

The main systematic uncertainties on the flux upper limits come from photon propagation in ice, absolute DOM
efficiency, and uncertainties in the Earth's density profile and
muon energy loss. For an $E^{-2}$ spectrum, the estimated total
uncertainty is about 16\% \citep{Jon}. They are included in the upper
limits calculations following the method of \citet{Conrad:2002kn} with
the modification described in \citet{Hill:2003jk}.

\section{Method}\label{sec:method}

The likelihood method used in this analysis was described in full detail and demonstrated in \citet{Braun:2008bg_Methods,Braun:2009wp}, and applied to the 40-string data in \citet{mike}. 
In the likelihood ratio method, the data are modeled as a combination of signal and background populations.
The probability density functions (PDFs) for signal
and background consist of three terms: a space term, an energy term and a time term. The first two are implemented in the same way as in \citet{Jon}.
For signal, the space term characterizes the clustering of event directions around the hypothesized source location (effectively, the point spread function for the reconstructed muon, since the interaction angle between the incoming neutrino and outgoing muon is subdominant at the energies of these data samples).
For background, the space term is simply estimated by time-scrambling of the real data.
The energy term for background is similarly a PDF built from the energy estimates of events in the real data (selected from a declination band similar to the declination of the source being searched for).  For signal, distinct energy PDFs are constructed for simulated events arising from a range of neutrino source spectra from $E^{-1}$ to $E^{-4}$.  The chief purpose of the energy term in this search is not to determine the spectrum of the source (if one were detected).  Rather, it is to enhance the detectability of a source if its spectrum is relatively hard (e.g. $E^{-2}$) by leveraging the difference in the energy distribution of the signal events compared to nearby background events, which are primarily atmospheric neutrinos with a soft ($\sim E^{-3.7}$) spectrum.

The third term in the PDF incorporates timing information.  For signal, a periodic emission with Gaussian time profile is assumed.  The period is fixed to that determined by photon observations, while the phase and duration of the neutrino emission are left as free parameters.  A Gaussian shape is used for the profile to provide a smooth function with the fewest assumptions about the exact time profile of the neutrino emission.  The time PDF for the $i$th event can thus be expressed as:

\begin{equation}
S^{\mathrm{time}}_{i} = \frac{1}{\sqrt{2\pi}\sigma_T} e^{ -\frac{ \left| \varphi_i-\varphi_0 \right|^2}{2\sigma_T^2} },
\end{equation}
where $\sigma_T$ is the width of the Gaussian, $\varphi_i$ is the
phase of the event and $\varphi_0$ is the phase of the peak of the neutrino emission. 
The fit parameters are $\sigma_T$ and $\varphi_0$.  For background, the time term is a flat function, because in the absence of detector biases the background events are randomly distributed in time. 

For each candidate source, the likelihood ratio analysis finds best-fit values for four parameters: the number of signal events, the spectral index of the signal, the peak phase of the signal and its duration.  An initial estimate of the significance is made by assuming the likelihood ratio follows a $\chi^2$ distribution and converting to a (pre-trial) p-value.  To ensure a robust estimate of the final significance, however, this assumption is not used, and a correction for the number of trials is also included.  For the final significance, the analysis is performed on time-scrambled data and the same catalog of sources.  The final post-trial p-value is given by the fraction of analyses which yield a smaller (pre-trial) p-value for any of the sources in the catalog.

\begin{figure}[htb]
  \centering
  \begin{tabular}{cc}
   \includegraphics[clip,width=3.2in]{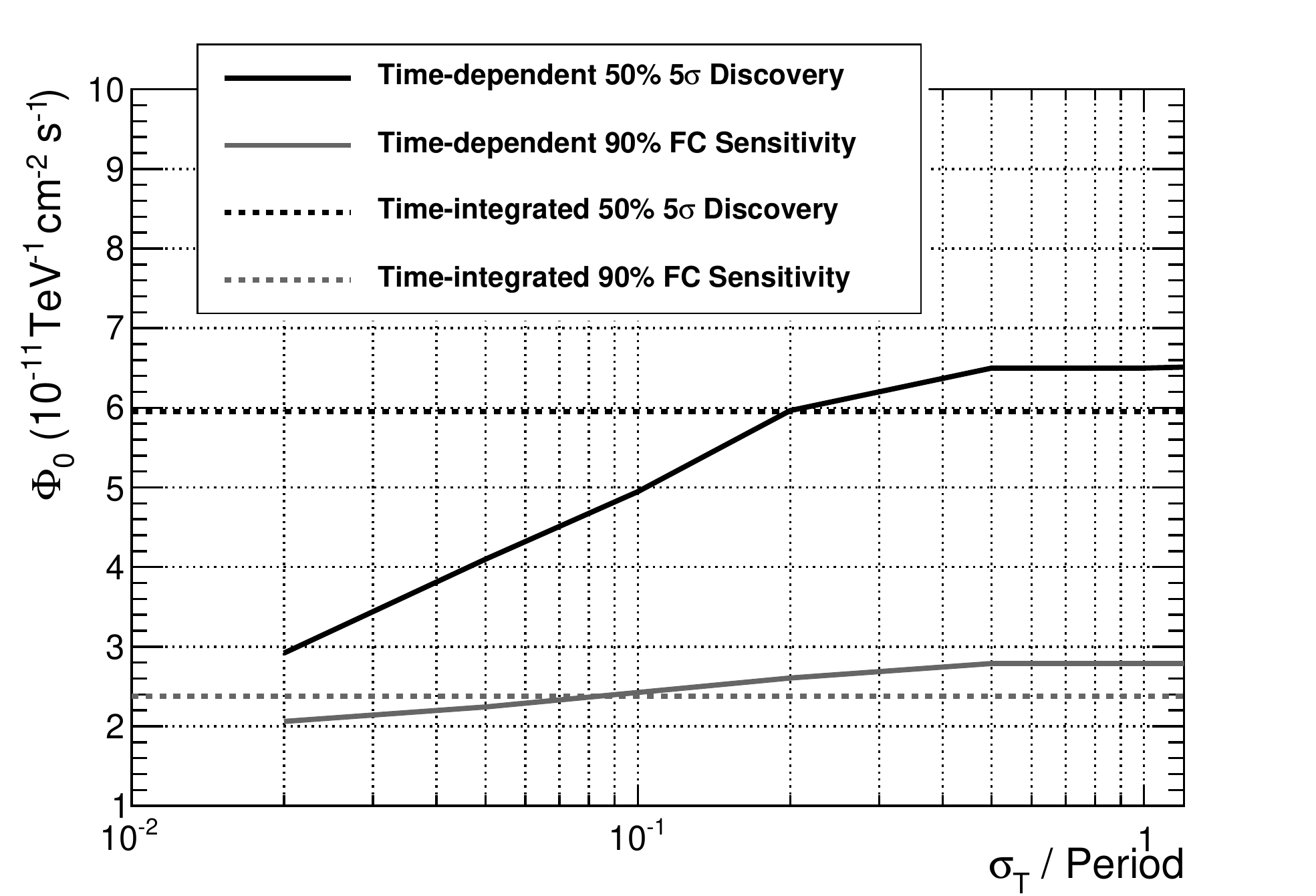} &
   \includegraphics[clip,width=3.2in]{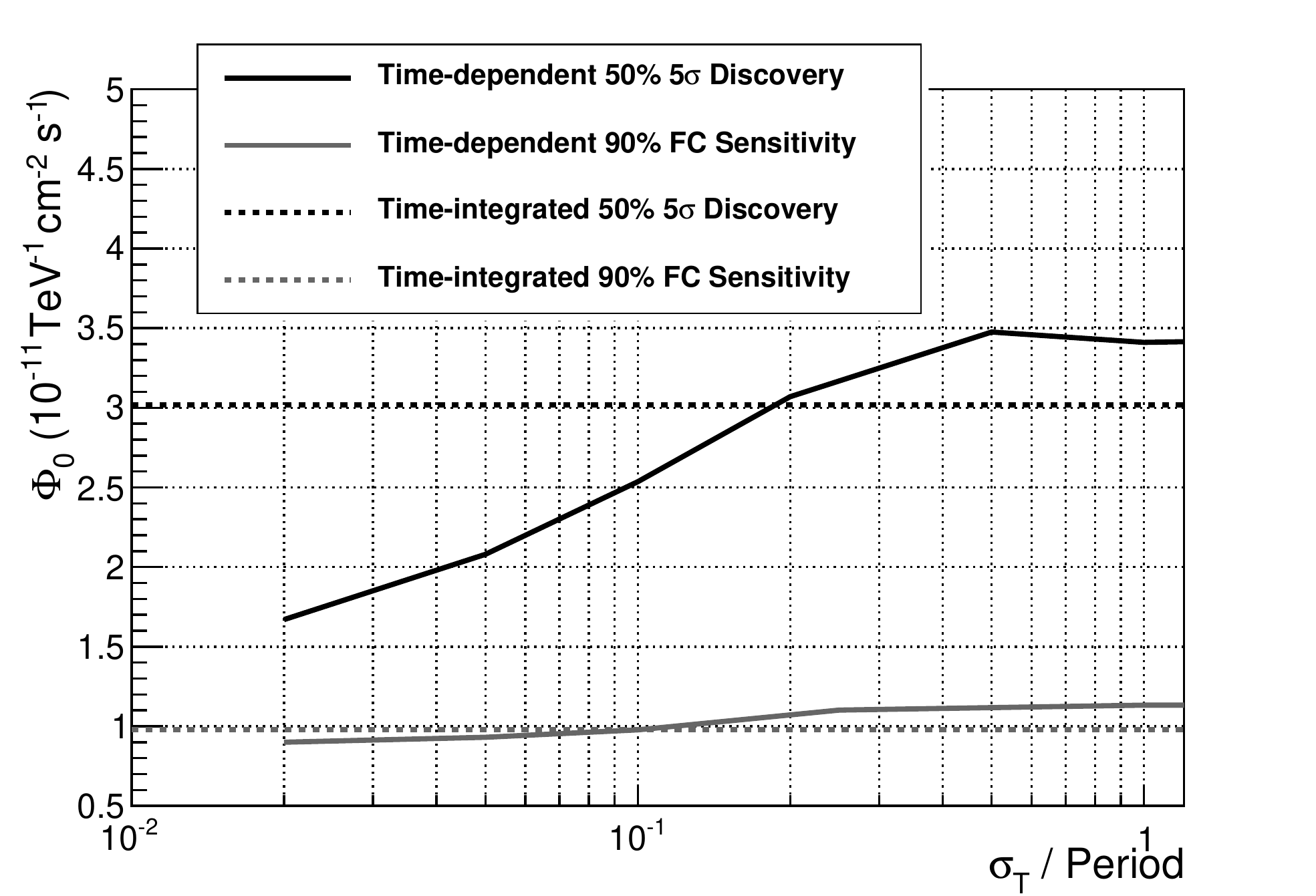} 
  \end{tabular}
  \caption{Discovery potentials (5$\sigma$ in 50\% trials, solid and dashed upper lines in the plot)
    and the average sensitivity at 90\% CL calculated with the prescription in \protect\citep{Feldman:1997qc} (solid and dashed lower lines) 
    for the periodic search applied to the 22-string data (left) and to the 40-string data (right). The y-axis shows the flux
    normalization $\Phi_0$ for a $E^{-2}$ simple power law
    spectrum emission from \lsi\ , i.e.~$\mathrm{d}\Phi/\mathrm{d}E =
    \Phi_0$\,$(E/\mathrm{TeV})^{-2}$. The x-axis shows the
    width $\sigma_T$ of the Gaussian emission normalized to
    its period $P = 26.4960\,$d. Shown are the results for the
    time-integrated searches \citep{Abbasi:2009iv, Jon} (dashed line) and for the periodic
    time-dependent search (solid line). The time-dependent search sensitivity and 
    discovery potential are expressed as the equivalent flux normalization of a steady source
    averaged over the period.}
  \label{fig:IC22_comp_sens}
\end{figure}

We calculate the sensitivity and median upper limit at 90\% confidence level using the method 
in \citet{Feldman:1997qc}. 
The discovery potential is the flux required to achieve a p-value less than 2.87$\times 10^{-7}$ (5$\sigma$ of the upper tail of a one-sided Gaussian)
 in 50\% of trials. It should be noted that the threshold significance to claim a discovery
in IceCube is set to 5$\sigma$.
Fig.~\ref{fig:IC22_comp_sens} shows the sensitivity and the discovery potential for the 
analysis,
together with the corresponding values from the
time-integrated search \citep{Abbasi:2009iv}.

Compared to the time-integrated analysis, searching for
periodicity in neutrino emission results in a better discovery
potential if the duration of the emission $\sigma_T$ is less than
about 20\% of the total period (see Fig.~\ref{fig:IC22_comp_sens}). As the
time-dependent search adds two additional degrees of freedom to the
analysis, the discovery potential is on the other hand 
roughly 10-15\% better using the time-integrated search if neutrinos
are actually emitted at a steady rate or over a large fraction of the
period. For both the 22-string and 40-string analyses, 
if the emission has a $\sigma_T$ of 1/50, 
the method requires about half as many events for discovery as the time-integrated search.

\section{Results}\label{sec:results}
The seven predefined sources, listed in
Table~\ref{tab:mqso}, were used for the initial search with the 22-string data from 2007-2008. 
The most significant outcome in this sample was for the source SS 433, with a pre-trial estimated p-value of 6\%.  In identical analyses of time-scrambled data, we find at least one of the seven tested sources to be more significant in 35\% of the analyses. 

The analysis was subsequently performed on the 40-string data from 2008-2009, which provided twice the sensitivity of the previous sample (see Fig.~\ref{fig:IC22_comp_sens}).
The most significant outcome in this sample was for Cygnus X-3. The pre-trial estimated
p-value of this source is 0.0019. 
To account for trials and for the fact that the likelihood ratio is not perfectly $\chi^2$ distributed, the analysis is performed again on time-scrambled data.
An equivalent or more significant outcome from any of the sources is found in 1.8\% of scrambled samples (expressed in Gaussian standard deviations, a $2.1\,\sigma$ excess), so the result is compatible with random fluctuations of the background.  The best-fit peak emission is found to be
at phase $\hat{\varphi}_0 = 0.82$, and $\hat{\sigma}_T = 0.02$. 
The best-fit number of
source events is $\hat{n}_s = 4.28$ and spectral index is $\hat{\gamma}_s = 3.75$. 
The full results of the analysis on each source (with time-dependent and time-integrated flux upper limits) are given in Table~\ref{tab:mqso}.

\begin{sidewaystable} 
 \begin{center}
   \scriptsize
   \begin{tabular}{@{}|c|c|c|c|c|c|>{\tiny}c|c|c|@{}}
     \tableline
     & Period & p-value &$T_0$ & $\hat{\varphi}_{o}$  & $\hat{\sigma}_{T}$ & & Time-Dependent UL & Time-Integrated UL \\
     Source & (days) & (pretrial) &(MJD) & (phase) & (period$^{-1}$) &  Reference & (TeV$^{-1}$cm$^{-2}$s$^{-1}$) & (TeV$^{-1}$cm$^{-2}$s$^{-1}$)\\
     \hline
     \cygxt\ &  0.199679 & 0.0019 & 54896.693 & 0.819 & 0.02 & \citep{Fermi12112009}
     & 3.01$\cdot10^{-11}$ & 6.64$\cdot10^{-12}$ \\
     \cygxo\ & 5.5929 & 0.080 & 41874.707 & 0.031 & 0.02 & \citep{Brocksopp:1998jw} 
     & 4.08$\cdot10^{-12}$ & 7.41$\cdot10^{-12}$ \\
     \lsi\ &  26.498    & 0.23 & 43366.775 & 0.916 & 0.02 & \citep{Acciari:2008hg} 
     & 1.82$\cdot10^{-11}$ & 9.78$\cdot10^{-12}$ \\
     \grs\  & 30.8      & 0.43  & 53945.7 & 0.502 & 0.045 & \citep{Neil:2006qa} 
     & 2.57$\cdot10^{-12}$ & 3.23$\cdot10^{-12}$ \\
     \ssf\ & 13.0821   & 0.35 & 50023.62 & 0.779 & 0.02 & \citep{Hillwig:2007us} 
     & 3.15$\cdot10^{-12}$ & 3.03$\cdot10^{-12}$ \\
     \xte\ &  0.169934 & 0.28 & 52287.9929 & 0.985 & 0.13 & \citep{McClintock:2003ns} 
     & 7.29$\cdot10^{-12}$ & 8.18$\cdot10^{-12}$ \\
     \gro\ & 0.21214   & 0.037 & 50274.4156 & 0.831 & 0.02 & \citep{Webb:2000xx} 
     & 1.46$\cdot10^{-11}$ & 6.89$\cdot10^{-12}$ \\
     \tableline
   \end{tabular}
   \caption{Results of the analysis with the IceCube 40-string data sample from 2008-2009.
    $T_0$ is the time of zero phase for the binary systems tested. 
    $\hat{\sigma}_{T}$ is the standard deviation of the best-fit Gaussian, as a fraction
     of the period of the binary system.
     The value for $\sigma_{T}$ is constrained to be 
     larger than 0.02, to prevent the method from isolating 
     a single event. We also include the
     reference used for the orbital information. The last columns
     give the 90\% CL upper limits on the normalization of the flux for
     an $E^{-2}$ neutrino spectrum, for the time-dependent
     and time-integrated hypotheses. The upper limits 
     also incorporate a 16\% systematic uncertainty.
   }
   \label{tab:mqso}
 \end{center}
\end{sidewaystable}

Table~\ref{mqso_dcomp} compares the 40-string time-integrated limits to the model predictions in \citet{Distefano:2002qw} for each source. 
The model predicts the neutrino flux based on the radiative luminosity associated to the jet from radio
observations in quiescent states and during flares the durations of which are specified in Tab.~4 in that paper. 
The figure shows limits for both the persistent and time-dependent cases for a time window similar to the observed flare but not coincident to it 
(since IceCube was not active at the time of radio observations noted in the paper).
For the persistent case of SS 433 the model predicts more than 100
events during the 40-string data taking period, a flux level which is excluded
by previous searches by the AMANDA detector \citep{Achterberg:2006vc}.
\citet{Distefano:2002qw} noted that for the specific case of SS 433, the model 
may be biased because the source is surrounded by the diffuse nebula W50, which 
can affect the estimate of the radio emission used in the model for this source.
For Cygnus X-3, the IceCube limits are near the prediction with the 40-string data.

\begin{table}
\centering{
\begin{tabular} {|l|l|c|c|c|}
\hline
Source       & Emission duration 
& \begin{sideways}Time-Integrated\end{sideways} \begin{sideways}90\% UL (events)\end{sideways}
& \begin{sideways}Model\end{sideways} \begin{sideways}Prediction (events)\end{sideways} 
& \begin{sideways}Time-Dependent\end{sideways} \begin{sideways}Sensitivity (events)\end{sideways} \\
\hline
Cygnus X3    & 3 days     & 5.1 & 3.0   & 2.9 \\
Cygnus X1    & Persistent & 3.3 & 1.9   & --- \\
LS I +61 303 & Periodic, 26\% duty cycle & 5.9 & 1.3 & 5.0 \\
GRS 1915+105 & 6 days     & 4.3 & 0.5   & 2.8 \\
SS 433       & Persistent & 3.3 & 220   & --- \\
GRO J0422+32 &  1 days    & 5.1 & 0.068 & 2.9 \\
             & 20 days    &     & 1.6   & 2.7 \\
\hline
\end{tabular}
}
\caption{The time-integrated upper limit (UL) at 90\% CL on the number of events is compared to the expected number of events for model predictions according to \protect\citet{Distefano:2002qw} for specific sources for the 40-string configuration. The neutrino energy range used to calculate the total number of events is $10^{11} - 5\times10^{14}$ eV, comparable to what was assumed in the model. 
For non-persistent but flaring sources, the parameters of the model were estimated for flares observed before IceCube construction. Hence the time-dependent sensitivities are calculated averaging
over a duration equal to the model flare during 40-string data
taking. 
\lsi\ is modeled as a periodically flaring source in a high state during 26\% of the orbit.}
    \label{mqso_dcomp}
    \end{table}

The main parameters on which the neutrino flux depends in this model are:
the fraction of jet kinetic energy converted to internal energy of electrons 
and magnetic field, $\eta_e$; the fraction of the jet luminosity carried by 
accelerated protons, $\eta_p$; and the fraction of proton energy in pions $f_\pi$, 
which strongly depends on the maximum energy to which protons can be accelerated.
We show as an example for the case of a 3-day burst of \cygxt\ how the parameters are constrained 
by our result. We assume equipartition between the magnetic fields and the 
electrons and the proton component ($\eta_p = \eta_e$) for setting a constraint on 
$f_\pi < 0.11$. If equipartition does not apply, we assume 
$f_\pi = f_{\pi,peak}$ as given in Table~2 in  \citet{Distefano:2002qw} (for \cygxt\ 
$f_{\pi,peak} = 0.12$) and constrain $\eta_p$ to be less than 92\% of
$\eta_e$. In deriving these limits we have assumed that the Lorentz
factor of the jet is well known from radio measurements, but in many
cases there is a large uncertainty on this parameter. Hence, 
our limits for the parameters of this model may have different
implications that we cannot disentangle: protons may not be dominant
in the jet, they may lose smaller energies into pion decay than the values
considered in \citet{Distefano:2002qw}, or the Lorentz factor is lower than the
value indicated in Table~1 in that paper.

\section{Conclusions}
The exploration of the GeV and TeV photon sky with the instruments on
board the Fermi spacecraft and the ground-based Cherenkov telescopes
has heralded the golden age of $\gamma$-ray astronomy. The
connection to neutrino astronomy is clear: high energy processes which 
cause the observed VHE emission can be responsible for the observed
high energy cosmic rays. This implies hadronic acceleration mechanisms
in astrophysical sources which can result in an observable neutrino
flux with giant neutrino telescopes like IceCube.

The available photon observations have made it possible to enhance the
sensitivity of searches for neutrino fluxes by incorporating
assumptions derived from the $\gamma$-ray data. One crucial
development has been the formulation of time-dependent neutrino flux
searches, postulating a connection between the time modulation of
the high energy emission and the possible neutrino flux. This
assumption has increased the sensitivity of these searches in
comparison to their time-averaged counterparts. 

This paper presented a search for neutrinos from objects with
periodic photon/broadband emission. Seven X-ray binaries in the
Northern Hemisphere were selected as candidate sources in
analyses of IceCube 22-string and 40-string data.
The most significant source in the catalog is \cygxt\ with a
1.8\% probability after trials ($2.1 \sigma$ excess).
Comparing the time-integrated limits for each source to
model predictions from \citet{Distefano:2002qw}, 
we show that our limits can constrain the
fraction of jet luminosity which is converted into pions and the ratio of
jet energy into relativistic leptons versus relativistic hadrons, under some assumptions.
For instance, for \cygxt\ and equipartition between electrons and protons, the 
fraction of proton energy in pions is limited to about 11\%.
All of the results in this paper
are compatible with a fluctuation of the background.

\section{Acknowledgments}
We thank D.~Guetta and E.~Waxman for helpful discussions on
neutrino flux prediction models.
We acknowledge the support from the following agencies:
U.S. National Science Foundation-Office of Polar Programs,
U.S. National Science Foundation-Physics Division,
University of Wisconsin Alumni Research Foundation,
the Grid Laboratory Of Wisconsin (GLOW) grid infrastructure at the University of Wisconsin - Madison, the Open Science Grid (OSG) grid infrastructure;
U.S. Department of Energy, and National Energy Research Scientific Computing Center,
the Louisiana Optical Network Initiative (LONI) grid computing resources;
National Science and Engineering Research Council of Canada;
Swedish Research Council,
Swedish Polar Research Secretariat,
Swedish National Infrastructure for Computing (SNIC),
and Knut and Alice Wallenberg Foundation, Sweden;
German Ministry for Education and Research (BMBF),
Deutsche Forschungsgemeinschaft (DFG),
Research Department of Plasmas with Complex Interactions (Bochum), Germany;
Fund for Scientific Research (FNRS-FWO),
FWO Odysseus programme,
Flanders Institute to encourage scientific and technological research in industry (IWT),
Belgian Federal Science Policy Office (Belspo);
University of Oxford, United Kingdom;
Marsden Fund, New Zealand;
Japan Society for Promotion of Science (JSPS);
the Swiss National Science Foundation (SNSF), Switzerland;
A.~Gro{\ss} acknowledges support by the EU Marie Curie OIF Program;
J.~P.~Rodrigues acknowledges support by the Capes Foundation, Ministry of Education of Brazil.




\begin{thebibliography}{44}
\expandafter\ifx\csname natexlab\endcsname\relax\def\natexlab#1{#1}\fi

\bibitem[{Abbasi {et~al.}(2009{\natexlab{a}})}]{Abbasi:2009iv}
Abbasi, R., {et~al.} 2009{\natexlab{a}}, \apj, 701, L47

\bibitem[{Abbasi {et~al.}(2009{\natexlab{b}})}]{IceCubeDAQ_:2008ym}
---. 2009{\natexlab{b}}, Nucl. Inst. \& Meth. in Phys. Res., A601, 294

\bibitem[{Abbasi {et~al.}(2010)}]{Abbasi:2010vc}
---. 2010, Nucl. Inst. \& Meth. in Phys. Res., A618, 139

\bibitem[{Abbasi {et~al.}(2011)}]{Jon}
---. 2011, {\apj}, 732, 18

\bibitem[{Abbasi {et~al.}(2012)}]{mike}
---. 2012, {\apj}, 744, 1

\bibitem[{Abdo {et~al.}(2009{\natexlab{a}})}]{Abdo:2009pw}
Abdo, A.~A., {et~al.} 2009{\natexlab{a}}, \apj, 701, L123

\bibitem[{Abdo {et~al.}(2009{\natexlab{b}})}]{Fermi12112009}
---. 2009{\natexlab{b}}, Science, 326, 1512

\bibitem[{Abdo {et~al.}(2011)}]{Abdo:2011dj}
---. 2011, \apj, 736, L11

\bibitem[{Acciari {et~al.}(2009)}]{Acciari:2009gg}
Acciari, V., {et~al.} 2009, 700, 1034

\bibitem[{Acciari {et~al.}(2008)}]{Acciari:2008hg}
Acciari, V.~A., {et~al.} 2008, \apj, 679, 1427

\bibitem[{Achterberg
  {et~al.}(2006)}]{IceCubeFirstYearPerformance_Achterberg:2006md}
Achterberg, A., {et~al.} 2006, Astropart. Phys., 26, 155

\bibitem[{Achterberg {et~al.}(2007)}]{Achterberg:2006vc}
---. 2007, Phys.Rev., D75, 102001

\bibitem[{Aharonian {et~al.}(2005{\natexlab{a}})}]{Aharonian:2005bv}
Aharonian, F., {et~al.} 2005{\natexlab{a}}, \aap, 442, 1

\bibitem[{Aharonian {et~al.}(2005{\natexlab{b}})}]{Aharonian:2005eb}
---. 2005{\natexlab{b}}, Science, 309, 746

\bibitem[{Aharonian {et~al.}(2007)}]{Aharonian:2007nh}
---. 2007, \aap, 469, L1

\bibitem[{Albert {et~al.}(2006)Albert, Aliu, Anderhub,
  {et~al.}}]{Albert23062006}
Albert, J., Aliu, E., Anderhub, H., {et~al.} 2006, Science, 312, 1771

\bibitem[{Albert {et~al.}(2007)}]{Albert:2007mf}
Albert, J., {et~al.} 2007, \apj, 665, L51

\bibitem[{Albert {et~al.}(2009)}]{Albert:2009lsi}
---. 2009, \apj, 693, 303

\bibitem[{Athar {et~al.}(2000)Athar, Je\ifmmode~\dot{z}\else \.{z}\fi{}abek, \&
  Yasuda}]{Athar}
Athar, H., Je\ifmmode~\dot{z}\else \.{z}\fi{}abek, M., \& Yasuda, O. 2000,
  Phys. Rev. D, 62, 103007

\bibitem[{Bongiorno {et~al.}(2011)Bongiorno, Falcone, Stroh, Holder, Skilton,
  Hinton, Gehrels, \& Grube}]{Bongiorno:2011vx}
Bongiorno, S.~D., Falcone, A.~D., Stroh, M., Holder, J., Skilton, J.~L.,
  Hinton, J.~A., Gehrels, N., \& Grube, J. 2011, ApJ, 737, L11

\bibitem[{Bosch-Ramon \& Khangulyan(2009)}]{BoschRamon:2008su}
Bosch-Ramon, V., \& Khangulyan, D. 2009, Int.J.Mod.Phys., D18, 347, * Brief
  entry *

\bibitem[{Braun {et~al.}(2008)}]{Braun:2008bg_Methods}
Braun, J., {et~al.} 2008, Astropart. Phys., 29, 299

\bibitem[{Braun {et~al.}(2010)}]{Braun:2009wp}
---. 2010, Astropart. Phys., 33, 175

\bibitem[{Brocksopp {et~al.}(1998)Brocksopp, Tarasov, Lyuty, \&
  Roche}]{Brocksopp:1998jw}
Brocksopp, C., Tarasov, A.~E., Lyuty, V.~M., \& Roche, P. 1998, \aap, 343, 861

\bibitem[{Conrad {et~al.}(2003)Conrad, Botner, Hallgren, \& Perez de~los
  Heros}]{Conrad:2002kn}
Conrad, J., Botner, O., Hallgren, A., \& Perez de~los Heros, C. 2003,
  Phys.Rev., D67, 012002

\bibitem[{Distefano {et~al.}(2002)Distefano, Guetta, Waxman, \&
  Levinson}]{Distefano:2002qw}
Distefano, C., Guetta, D., Waxman, E., \& Levinson, A. 2002, \apj, 575, 378

\bibitem[{Feldman \& Cousins(1998)}]{Feldman:1997qc}
Feldman, G.~J., \& Cousins, R.~D. 1998, \prd, 57, 3873

\bibitem[{Gregory(2002)}]{0004-637X-575-1-427}
Gregory, P.~C. 2002, \apj, 575, 427

\bibitem[{{Gregory} \& {Taylor}(1978)}]{1978Natur.272..704G}
{Gregory}, P.~C., \& {Taylor}, A.~R. 1978, \nat, 272, 704

\bibitem[{{Harrison} {et~al.}(2000){Harrison}, {Ray}, {Leahy}, {Waltman}, \&
  {Pooley}}]{2000ApJ...528..454H}
{Harrison}, F.~A., {Ray}, P.~S., {Leahy}, D.~A., {Waltman}, E.~B., \& {Pooley},
  G.~G. 2000, \apj, 528, 454

\bibitem[{Hill(2003)}]{Hill:2003jk}
Hill, G.~C. 2003, Phys.Rev., D67, 118101

\bibitem[{Hillwig \& Gies(2008)}]{Hillwig:2007us}
Hillwig, T.~C., \& Gies, D.~R. 2008, \apj, 679, 1427

\bibitem[{Li {et~al.}(2011)Li, Torres, Zhang, Chen, Hadasch, Ray, Kretschmar,
  Rea, \& Wang}]{0004-637X-733-2-89}
Li, J., {et~al.} 2011, \apj, 733, 89

\bibitem[{McClintock {et~al.}(2003)}]{McClintock:2003ns}
McClintock, J.~E., {et~al.} 2003, \apj, 593, 435

\bibitem[{Neil {et~al.}(2007)Neil, Bailyn, \& Cobb}]{Neil:2006qa}
Neil, E.~T., Bailyn, C.~D., \& Cobb, B.~E. 2007, \apj, 657, 409

\bibitem[{Paredes(2011)}]{Paredes:2011vh}
Paredes, J.~M. 2011, Nuovo Cimento C, 034, 167

\bibitem[{{Paredes} {et~al.}(2000){Paredes}, {Mart{\'{\i}}}, {Rib{\'o}}, \&
  {Massi}}]{2000Sci...288.2340P}
{Paredes}, J.~M., {Mart{\'{\i}}}, J., {Rib{\'o}}, M., \& {Massi}, M. 2000,
  Science, 288, 2340

\bibitem[{{Sidoli} {et~al.}(2006){Sidoli}, {Pellizzoni}, {Vercellone},
  {Moroni}, {Mereghetti}, \& {Tavani}}]{2006AnA...459..901S}
{Sidoli}, L., {Pellizzoni}, A., {Vercellone}, S., {Moroni}, M., {Mereghetti},
  S., \& {Tavani}, M. 2006, \aap, 459, 901

\bibitem[{{Tavani} \& {Arons}(1997)}]{1997ApJ...477..439T}
{Tavani}, M., \& {Arons}, J. 1997, \apj, 477, 439

\bibitem[{Tilav {et~al.}(2009)}]{Tilav:2010hj}
Tilav, S., {et~al.} 2009, in Proc. of the 31st Int. Cosmic Ray Conf, Lodz,
  Poland, {\it Atmospheric Variations as observed by IceCube}, arXiv:1001.0776

\bibitem[{Torres {et~al.}(2010)Torres, Zhang, Li, Rea, Caliandro, Hadasch,
  Chen, Wang, \& Ray}]{2041-8205-719-1-L104}
Torres, D.~F., {et~al.} 2010, \apjl, 719, L104

\bibitem[{Webb {et~al.}(2000)Webb, Naylor, Ioannou, Charles, \&
  Shahbaz}]{Webb:2000xx}
Webb, N.~A., Naylor, T., Ioannou, Z., Charles, P.~A., \& Shahbaz, T. 2000,
  \mnras, 317, 528

\bibitem[{Zhang {et~al.}(2010)Zhang, Torres, Li, Chen, Rea, \&
  Wang}]{MNR:MNR17152}
Zhang, S., Torres, D.~F., Li, J., Chen, Y.~P., Rea, N., \& Wang, J.~M. 2010,
  \mnras, 408, 642

\bibitem[{{Zi{\'o}{\l}kowski}(2005)}]{2005MNRAS.358..851Z}
{Zi{\'o}{\l}kowski}, J. 2005, \mnras, 358, 851

\end{thebibliography}
\end{document}